\documentclass[aps,prb,twocolumn,groupedaddress]{revtex4-1}
\usepackage[pdftex]{graphicx}
\usepackage{bm}
\usepackage{etoolbox}
\usepackage[usenames,dvipsnames]{color}
\usepackage{mathtools}
\usepackage{amsmath}%
\usepackage{amsfonts}%
\usepackage{amssymb}
\usepackage{graphicx}
\usepackage[breaklinks=true,colorlinks=true,linkcolor=blue,urlcolor=blue,citecolor=blue]{hyperref}
\usepackage{bbm}

\DeclarePairedDelimiterX\braket[2]{\langle}{\rangle}{#1 \delimsize\vert #2}

\newcommand{\bg}{ \begin{gather} }

\newcommand{\eg}{\end{gather}}
\newcommand{\be}{ \begin{equation} }
\newcommand{\ee}{\end{equation}}
\newcommand{\bea}{ \begin{eqnarray} }
\newcommand{\eea}{\end{eqnarray}}

\renewcommand{\Im}{\mathop{\rm Im}}

\begin{document}

\title{Electron-phonon cooling power in Anderson insulators}
\author{M. V. Feigel'man}
\affiliation{ L. D. Landau Institute for Theoretical Physics, Kosygin str.2, Moscow 119334, Russia}
\affiliation{Skolkovo Institute of Science and Technology, 143026 Skolkovo, Russia}
\author{V. E. Kravtsov}
\affiliation{ Abdus Salam International Center for Theoretical Physics, Trieste, Italy}
\affiliation{ L. D. Landau Institute for Theoretical Physics, Kosygin str.2, Moscow 119334, Russia}

\begin{abstract}
We present a microscopic theory for electron-phonon energy exchange in 
Anderson insulators at low temperatures.
The major contribution to the cooling power $J_{e-ph}(T_{el})$ as a function of electron temperature $T_{el}$
is shown to be  directly related to the correlation function of the local density of electron states $K(\omega)$.
In Anderson insulators not far from localization transition, correlation function  $K(\omega) $ is  enhanced 
 at small $\omega$  by  wavefunction's multifractality and by the presence of  Mott's resonant pairs of states.
The theory we develop explains a huge enhancement of the cooling power observed in insulating Indium Oxide films 
as compared to predictions of the theory previously developed for disordered metals.  
\end{abstract}

\maketitle
\section{Introduction.}
 
Energy exchange between electrons and phonons is  crucial to many physical 
properties of Anderson insulators at low temperatures: it 
determines relatively slow rate of thermal
equilibration.   Surprizingly, no theory of such processes seems 
to be availalble. On the contrary,
theory of electron-phonon inelastic coupling in disordered metals 
is known for a very long 
time~\cite{pippard,akhiezer,schmid}.   

Experimentally, one of the most sensitive method to study electron-phonon 
cooling rate is based on the results of 
Ref.\cite{shahar,Ovadia-Sacepe-Shahar}  where  striking jumps by several 
orders of magnitude  in current-voltage characteristics were observed at 
low temperatures in insulating  Indium Oxide films. Similar effects were 
also observed in other insulating systems \cite{sanquer, baturina}. These 
jumps in resistance are the signatures of thermal bi-stability at weak 
electron-phonon 
coupling   which can be analyzed using the  balance between the Joule 
heat production 
and the electron  cooling power
~\cite{AKLA},  and the temperature dependence of electron-phonon cooling 
rate can 
be experimentally obtained ~\cite{Ovadia-Sacepe-Shahar}. 
The out-cooling rate at 
low temperatures of electron system $T=T_{{\rm el}}$ appeared to be 
$J(T)= A\, T^{\beta}$, where $\beta\approx 6$ agreed well with the 
theory of electron-phonon cooling  presented in \cite{AKLA}.   

\begin{figure}
\center{\includegraphics[width=80mm]{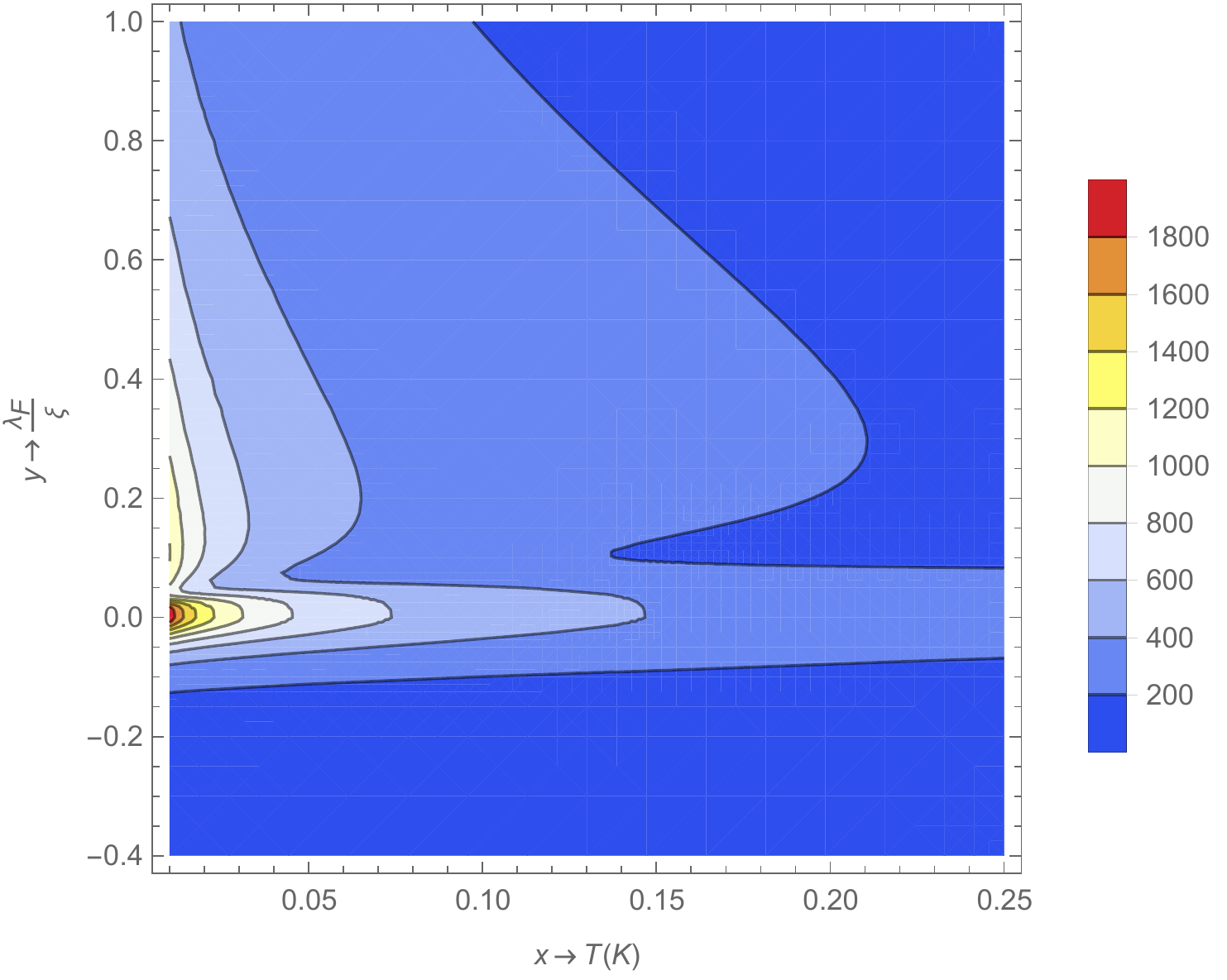}}
\caption{ Enhancement factor for the cooling rate in the Anderson insulator ($y>0$) 
compared to the dirty metal ($y<0)$ as a function 
of temperature and proximity to the Anderson localization transition at $n=n_{c}$
(parametrized by the 
 ratio of the Fermi wavelength $\lambda_{F}$ and the localization ($\xi>0$) 
or correlation ($\xi<0$) length, 
with $|\xi|/\lambda_{F}\sim(1-n/n_{c})^{-\nu}$, $\nu\approx 1.6$). 
Notice a persistent 
character of enhancement in insulator at low temperatures even 
far from the Anderson transition $\lambda_{F}/\xi=0$. This enhancement is caused 
by the pairs of Mott's resonant states with multifractal 
structure inside localization volume, while the enhancement 
close to the Anderson transition both in metal and in insulator
 is caused by multifractality alone. }
\label{Fig:enhance}
\end{figure}

The problem with this result is that the experimentally observed pre-factor 
$A$ is 2-3 {\it orders of magnitude} larger 
than  the one predicted by the theory of electron-phonon cooling in strongly 
disordered metals employed in \cite{AKLA}. At first glance it is also strange 
that in an insulator the temperature dependence of the cooling is a power-law, 
while the temperature dependence of resistance is exponential or 
stretch exponential. 
However, the most surprising fact is that the theory of electron-phonon cooling in 
Anderson insulators is essentially missing, despite so much effort invested in 
studying hopping conductivity. 

In this paper we present the theory of electron-phonon cooling in insulators 
at proximity  to the Anderson 
localization transition when  the momentum relaxation rate $\ell$ is of the order 
of the  Fermi  wavelength $\lambda_{F}=2\pi/k_{F}$,
and the effects of multifractality \cite{CueKrav,MFSC} are significant. 
We show that the temperature dependence of the
 cooling rate at low temperatures is indeed a power-law, since 
the energy exchange between electron and phonon systems is 
{\it local} and  does not involve  electron transport in space. 
Therefore it is natural that the additional  factor 
characterizing electron cooling
in Anderson  insulators obtained in this paper is given 
by the properly normalized correlation function $K(\omega)$ of the
{\it local} density of states.  
This correlation function is enhanced due to multifractality of electron wave 
functions \cite{CueKrav, MFSC}, which  results in an increasing cooling 
rate both in 
metals and in insulators close 
to the Anderson transition (see Fig.\ref{Fig:enhance}).  
Another mechanism of enhancement of the cooling rate (also described by the same 
correlation function $K(\omega)$) is typical to insulators and is related with  the 
Mott's pairs of resonant states.
It is similar to the logarithmic enhancement of the frequency-dependent 
conductivity in Anderson insulator~\cite{Mott, Berezinskii} and efficient 
at low temperatures. It is because of this effect, enhanced by multifractality, 
that the enhancement 
factor shown 
in  Fig.\ref{Fig:enhance} is drastically asymmetric on both sides of the Anderson transition. 

At the values of the parameters   typical to amorphous Indium Oxide films used in 
Ref.\cite{shahar, Ovadia-Sacepe-Shahar}, 
the total enhancement 
factor may be as large as  500-800 in the range of electron temperatures 
$20-100\,mK$ and it decreases very slowly
as the system is driven deeper into the insulating phase (see Fig\ref{Fig:enhance}).  
Moreover, the temperature dependence of the enhancement factor is logarithmic, 
which makes the effective power $\beta$ in the out-cooling rate 
only slightly modified compared to the case of dirty metal \cite{fn1}.
This  makes our theory a very plausible explanation of enhancement of the 
pre-factor $A$ in the   
cooling rate in the experiments \cite{Ovadia-Sacepe-Shahar}.

However, the results of this paper are much more general.   
They are based on universal properties of random electron wave functions in the 
{\it multifractal insulator} \cite{CueKrav, MFSC} and are independent of 
a particular system as well of the presence or absence of superconductivity in it.

The paper is organized as follows. In Sec.II we present a general expression for 
the out-cooling rate in terms if exact electron wave functions in the 
presence of strong disorder. In Sec.III and Appendix B we show that the simple 
random-phase approximation for electron wave functions employed in 
the theory of Sec.II reproduces all the known results for the electron-phonon 
cooling obtained earlier using the impurity diagrammatic technique. In Sec.III we 
modify this random-phase approximation by introducing a non-trivial envelope 
of oscillating wave functions which accounts for the 
 effects of multifractality and localization.The main result of this section is
that the cooling rate is determined by the local density of states correlation function
$K(\omega)$. In Sec.IV we review known properties of function $K(\omega)$, 
in particular the signatures of multifractality and the effect of Mott's resonant 
pair on it. 
In Sections V and VI we compute the enhancement factor for the cooling rate due 
to these effects for the transverse and longitudinal phonons, respectively. 
In Conclusion we formulate the main results of the paper and discuss 
their implications for low-temperature experiments
in Anderson insulators close to localization transitions.

\section{ General expression for the cooling rate.}
The out-cooling rate $J(T)$ is expressed~\cite{Shtyk} in terms of the phonon 
attenuation rate $\tau_{\rm ph}^{-1}$ due to electron phonon interaction:
\be\label{J-tau}
J(T)=\int_{0}^{\infty}d\omega\,\omega\,\nu_{ph}(\omega)\,
\frac{B_{ph}(\omega)}{\tau_{ph}(\omega)},
\ee
where $B_{\rm ph}(\omega) =\frac{1}{2}(\coth(\omega/2T)-1)$ is 
the phonon energy distribution function, and $\nu_{\rm ph}=
\omega^{2}/(2\pi^{2}v_{s}^{3})$ is the 3d phonon density of states. 
The phonon attenuation rate and the sound velocity  are different 
for transverse (t) and longitudinal (l) modes, and the total cooling 
rate $J_{{\rm tot}}(T)=J_{{\rm l}}(T) + 2 J_{{\rm t}}(T)$, each of 
the contributions being described by (\ref{J-tau}) with the 
corresponding $\tau_{ph}^{(t,l)}$ and sound velocities $v_{s}^{(t,l)}$.

Thus the primary object of interest is the phonon attenuation rate:
\be\label{tau-Sigma}
\frac{1}{\tau_{ph}}=\frac{1}{2\rho_{i}\,\omega}\,\Im(\Sigma_{\omega}^{R}-
\Sigma_{\omega}^{A}),
\ee
where $\rho_{i}$ is the lattice mass density,  and $\Sigma^{R(A)}=
\hat{{\cal D}}\,\Pi^{R(A)}_{RPA}\,\hat{{\cal D}}$ is the (retarded or advanced) 
phonon self energy, given by a proper action of the gradient vertex operators 
$\hat{{\cal D}}$ 
on the RPA polarization bubble $\Pi^{R(A)}_{RPA}$.
 
In order to take the localized nature of electron wave functions into 
account we express the   phonon attenuation rate in terms of the exact electron 
eigenfunctions $\psi_{n}({\bf r})$ and eigenvalues $E_{n}$. To this end we use the  reference frame moving locally together with the lattice \cite{schmid, Shtyk}.
In this frame the electron-phonon Hamiltonian takes the form~\cite{Shtyk}:
\bea\label{e-ph-Ham}
&&H_{{\rm e-ph}}=-\sum_{{\bf p},{\bf q}}p_{\alpha}\,(v_{\beta}\nabla_{\beta}
\,u_{\alpha})_{{\bf q}}\,\Psi^{\dagger}_{{\bf p}}\Psi_{{\bf (p+q)}}\nonumber 
\\ &&=\frac{1}{m}\int d^{d}{\bf r}\,[\nabla_{\beta}\,u_{\alpha}({\bf r})]\,
\partial_{\alpha}\partial'_{\beta}\;\Psi^{\dagger} 
(\bf r)\,\Psi ({\bf r'})|_{{\bf r}={\bf r'}},
\eea
where  $p_{\alpha}=-i\nabla_{\alpha}$, $v_{\beta}=p_{\beta}/m$ is the 
electron velocity operator, $\Psi_{{\bf p}}$ and $\Psi_{{\bf p+q}}$ are 
Fourier components of  the Fermionic operators $\Psi^{\dagger}({\bf r})$ 
and $\Psi({\bf r'})$,  $m$ is the electron mass and $u_{\alpha}$ is the 
phonon-induced local  shift of the lattice in the laboratory frame. The 
Greek symbols  $\alpha, \beta$ in Eq.(\ref{e-ph-Ham}) and throughout the 
paper are the components of 3D vectors, the summation over repeated indexes 
being assumed. This Hamiltonian should be supplemented by the standard  electron  
interaction with an impurity potential and the electron kinetic energy.
The advantage of the co-moving frame is that the cross-terms with electron-
phonon-impurity interaction do not appear,
which makes calculations much simpler.

This interaction is screened by Coulomb interaction $V$. In the RPA 
approximation the screened phonon self-energy   is given by:
\be\label{screened}
\Sigma=\hat{{\cal D}}\Pi \hat{{\cal D}}+\hat{{\cal D}}\Pi\;\frac{V}
{1-\Pi V}\;\Pi \hat{{\cal D}},
\ee
where
\be
\hat{{\cal D}} \Pi=\frac{1}{m}\,[\nabla_{\beta}\,u_{\alpha}({\bf r})]\,
\partial_{\alpha}\partial'_{\beta}\,\Pi({\bf r},{\bf r'};{\bf r}_{1},{\bf r'}_{1})|_{{\bf r}={\bf r'},{\bf r}_{1}={\bf r}_{1}'},
\ee
\be
\Pi \hat{{\cal D}}= \frac{1}{m}\,[\nabla_{\gamma}\,u_{\delta}({\bf r}_{1})]\,
\partial_{1,\gamma}\partial'_{1,\delta}\,\Pi({\bf r},{\bf r'};{\bf r}_{1},
{\bf r'}_{1})|_{{\bf r}={\bf r'},{\bf r}_{1}={\bf r}_{1}'},
\ee
and $\Pi$ is the bare polarization bubble in which all effects of disorder are 
included but interaction is not.

Note that in the second term in Eq.(\ref{screened}) the fast momenta 
corresponding to the left vertex of the leftmost  $\Pi$ is completely 
decoupled from the fast momenta corresponding to the right vertex of the 
rightmost $\Pi$. As the result the second term in Eq.(\ref{screened})  is 
proportional to $k_{F}^{4}\,\delta_{\alpha\beta}\,\delta_{\gamma\delta}$ 
and thus its
contribution vanishes   for transverse phonons. This is not the case for 
the first term in Eq.(\ref{screened}) at distances $|{\bf r}-{\bf r}_{1}|<\ell$, 
where $\ell$ is the mean free path. 

In what follows we first consider the effect of the first term in 
Eq.(\ref{screened}).
Using (\ref{screened}),(\ref{e-ph-Ham}) and (\ref{tau-Sigma}) one can 
express the corresponding contribution to $\tau_{ph}$ as follows 
(see Appendix for details of derivation):
\begin{equation}\label{tau-gen}
\frac{1}{\tau_{{\rm ph}}^{(1)}} =\pi\,\frac{q_{\beta}q_{\delta}}{m^{2}}
\,e_{\alpha}e_{\gamma}\,\frac{1}{\rho_{i} }\,   
\int d^{d}{\bf R} \,e^{i{\bf q}{\bf R }}\,   K_{\alpha\beta\gamma\delta} 
({\bf R}, \omega),
\end{equation} 
where $e_{\alpha}$ is the $\alpha$ component the unit vector of phonon 
polarization, $q_{\alpha}$ is the component of the phonon wave vector 
${\bf q}$ with $|{\bf q}|=q=\omega/v_{s}$,    ${\cal V}$ is the volume,   
and the function 
$K_{\alpha\beta\gamma\delta} ({\bf R} ; \omega ) $  is defined as
\begin{widetext}
\begin{equation}
\label{K}
K_{\alpha\beta\gamma\delta} ({\bf R};\omega )= \left\langle\sum_{nm} 
(\partial_{\alpha}\psi^{*}_{m}({\bf r}))\,(\partial_{\beta}\psi_{n}({\bf r}))\,
(\partial_{\gamma}\psi^{*}_{n}({\bf r'}))\,(\partial_{\delta}\psi_{m}({\bf r'}))
\,\delta(E-E_{n})\,\delta(E'-E_{m})\right\rangle.
\end{equation}
 
\end{widetext}
In (\ref{K}) we denote disorder averaging by $\langle .\rangle$. After such an 
averaging  $K_{\alpha\beta\gamma\delta} ({\bf R},\varepsilon)$ becomes a function 
of ${\bf r}-{\bf r'}={\bf R}$ and $E-E'=\varepsilon$ in the bulk of a sample and 
the spectrum.  

\section{ Effects of localization and multifractality.}
To further proceed we employ the following ansatz for the electron wave functions:
\be\label{psi-ansatz}
\psi_{n}({\bf r})=\int \frac{d\Omega_{{\bf s}}}{4\pi}\,a^{(n)}_{{\bf s}}({\bf r})
\,e^{ik_{F}\,{\bf s}\,{\bf r}},
\ee   
where $|{\bf s}|=1$ and $a_{{\bf s}}^{(n)}({\bf r})$ is a Gaussian random variable 
with zero mean and the correlation function: 
\be\label{a-corr}
\langle a^{(n)}_{{\bf s}}({\bf r})\,a^{(m)}_{{\bf s'}}({\bf r'})\rangle= 
\delta_{nm} \,\delta_{{\bf s},{\bf s'}}\,e^{-\frac{|{\bf r}-{\bf r'}|}{2\ell}}\;
\phi_{n}({\bf r})\phi_{m}({\bf r'}).
\ee
Eqs.(\ref{psi-ansatz}),(\ref{a-corr}) are essentially a generalization of the 
semi-classical Berry' ansatz \cite{Berry} for the case of localization and 
multifractality.
The exponential factor with the momentum relaxation length $\ell$  in 
Eq.(\ref{a-corr}) accounts for the fast randomization
of wave-function phases due to elastic scattering, while  positive functions 
$\phi_{n}({\bf r})$ describe  normalized (and smooth at a scale $\ell$ ) \, 
{\it envelopes} of the wave functions, averaged over fast de Broglie oscillations:
\be\label{phi}
[\phi_{n}({\bf r})]^{2}=\overline{\psi_{n}^{2}({\bf r})}\;{\cal V}.
\ee
Such an envelopes  $\phi_{n}({\bf r})$ are equal to 1 in the semi-classical 
Berry's approximation $k_{F}\ell\gg 1$ when both localization and multifractality 
effects are absent  and wave functions are {\it ergodic}.  At $k_{F}\ell\sim 1$ 
when multifractality and/or localization is present, these envelope functions 
have a non-trivial shape which depends on the index $n$ and on the realization 
of disorder. Thus the averaging in (\ref{a-corr}) is incomplete. It involves only 
the random phase averaging and assumes subsequent disorder averaging of the 
amplitude.  Possibility to separate nearly universal
fast wave-functions oscillations from the slow envelope that contains information 
about
multifractal behavior was discussed in a different way in 
Ref.~\cite{MirlinFyodorov1997}. This idea has been successfully exploited  
in Ref.\cite{RRG} in numerical computation of the multifractal spectrum 
$f(\alpha)$ in order to sort out  the effect of nodes which dominates 
distribution of small eigenfunction amplitudes.

It is shown in the  Appendix A, that  plugging (\ref{psi-ansatz}) and 
(\ref{a-corr}) with $\phi_{n}({\bf r})=1$ in (\ref{tau-gen}) one exactly  
reproduces  at $q\ell\ll 1$ an   expression for  $\tau_{{\rm ph}}^{(t)}$ 
obtained 
earlier for diffusive metals~\cite{schmid, Reyzer, KrYud}: 
\be\label{old-tau}
\frac{1}{\tau_{{\rm ph}}^{(t)}}=\frac{q^{2}\,k_{F}^{4}\ell}{30\pi^{2}\,
\rho_{i}}=\frac{q^{2}}{10}\frac{k_{F}\ell}{\rho_{i}}\,n_{{\rm e}},
\ee
where $n_{{\rm e}}$ is the total (two-spin) electron density. The 
corresponding result for $J_{t}(T)$ is:
\be\label{old-J}
J_{t}(T)=\frac{4\pi^{4}}{630}\frac{(k_{F}\ell)n_{{\rm e}}}{\rho_{i}\,
[v_{s}^{(t)}]^{5}}\,T^{6}.
\ee
Taking now into account $\phi_{n}({\bf r})\neq 1$ in (\ref{a-corr})  
 one obtains for  (\ref{K}),(\ref{tau-gen}),(\ref{psi-ansatz}) the 
following expression for $\tau^{(1)}_{{\rm ph}}$:
\begin{widetext}
\be\label{tau-loc}
\frac{1}{\tau^{(1)}_{{\rm ph}}}=\pi\,\nu_{0}^{2}\,k_{F}^{4}\,
\frac{q_{\beta}q_{\delta}}{m^{2}\rho_{i}}\,e_{\alpha}e_{\gamma}\,
\int d^{3}{\bf R}\,e^{i{\bf q}{\bf R}}\,e^{-|{\bf R}|/\ell}\int
\frac{d\Omega_{{\bf s}}}{4\pi}\int\frac{d\Omega_{{\bf s'}}}{4\pi}
\,s_{\alpha}s_{\delta}s'_{\beta}s'_{\gamma}\,e^{-ik_{F}\,({\bf s}-{\bf s'})
{\bf R}}\,  K(\omega;{\bf R}),
\ee
 where $\nu_{0}$ is the mean density of states at the Fermi level, 
$\Delta=(\nu_{0}{\cal V})^{-1}$ is the mean level spacing in an entire 
volume ${\cal V}$ and
$K(\omega;{\bf R})= \Delta^{2} \left\langle \sum_{n,m}\phi_{n}({\bf r}) 
\,\phi_{m} ({\bf r'})\,\phi_{m}({\bf r})\,\phi_{n}({\bf r'})\;\delta(E-E_{n})
\delta(E+\omega-E_{m})\right\rangle$.
\end{widetext}
As the exponential factor $e^{-|{\bf R}|/\ell}$ makes the main domain of 
integration over ${\bf R}$   in (\ref{tau-loc}) to be  $|{\bf R}|\lesssim \ell$ 
and because of the smooth behavior of the envelope functions $\phi_{n}({\bf r})$ 
at such scale, one can replace $K(\omega;{\bf R})\rightarrow K(\omega,0)\equiv 
K(\omega)$.  Then after angular integration over unit vectors ${\bf s},{\bf s'}$ 
and integration over $R$ in (\ref{tau-loc}), 
 one obtains in the limit $|{\bf q}|\ell\ll 1$:
\be\label{tau-fin}
\frac{1}{\tau^{(1)}_{{\rm ph}}}=\frac{2\pi^{2}}{15}\frac{\nu_{0}^{2}k_{F}^{2}\ell}
{m^{2}\,\rho_{i}}\,(3q_{||}^{2}+q_{\perp}^{2})\;K(\omega),
\ee
where $q_{||}$ and $q_{\perp}$ are the longitudinal and the transverse components 
of the phonon wave vector and
\be\label{K-omega}
K(\omega)= \Delta^{2} \left\langle \sum_{n,m}\phi_{n}^{2}({\bf r}) \,\phi^{2}_{m}
 ({\bf r'}) \;\delta(E-E_{n})\delta(E+\omega-E_{m})\right\rangle
\ee
is the local density-of-states correlation function studied in
 Refs.~\cite{CueKrav,MFSC}.

For transverse phonons ($q_{||}=0$) Eq.(\ref{tau-fin}) gives the total 
phonon attenuation rate.
It is proportional to
the properly normalized electron local density of states correlation 
function $K(\omega)$  which is course-grained 
at a scale $\ell$. 
All the effects of localization and/or multifractality are encoded in this 
correlation function,  while  the effects of fast randomization of wave 
function phases by impurity scattering   are taken into account by averaging 
over momentum
directions $\mathbf{s},\mathbf{s'}$ in Eq.(\ref{tau-loc}).

Equations  (\ref{tau-fin}),(\ref{K-omega}) are the main result of our paper. 
Strictly speaking it is  valid in $2+\epsilon$ ($\epsilon\ll 1$) dimensions 
where the scale separation  $k_{F}\gg \ell^{-1}\gg \xi^{-1}$ holds even in 
insulator close to the Anderson transition where the localization length 
$\xi$ is large. As customary, we extend this result (with the accuracy up to 
a factor of order 1) for 3D samples and thick films with $k_{F}\ell\sim 1$. 
 
\section{The function $K(\omega)$ close to localization transition.}
The behavior of the correlation function $K(\omega)$ was studied in detail in 
Ref.\cite{CueKrav, MFSC}.   It was shown that for $E_{0}\gg\omega\gg\delta_{\xi}$,
 where $\delta_{\xi}=(\nu_{0}\xi^{3})^{-1}$ is the level spacing in the volume
characterized by the correlation/localization length $\xi$, 
and $E_{0}$ is of the order of total bandwidth of conduction band, the
 effects of multifractality  lead to the power-law enhancement of 
$K(\omega)=(E_{0}/\omega)^{\gamma}$, where $\gamma=1-d_{2}/3\approx 0.59$ 
is determined by the fractal dimension $d_{2}\approx 1.24\pm 0.03$ \cite{Roemer}. 
This effect is due to the non-ergodicity of wave functions which do not occupy 
all the available volume causing the enhancement of their amplitude by 
normalization. Furthermore, the support sets of different wave functions are 
strongly correlated thus giving rise to enhancement of the overlap function 
$K(\omega)$.  
Albeit  analysis  in \cite{CueKrav} concerned the case of non-interacting 
electrons, the subsequent study \cite{KrAminiMull} has shown that localization 
transition and multifractality survive almost unchanged when Coulomb interaction 
is taken into account.

As $\omega$ decreases below $\delta_{\xi}$ the behavior of $K(\omega)$ starts 
to depend on whether the system is insulating or 
metallic. In the latter case $K(\omega)\sim (E_{0}/\delta_{\xi})^{\gamma}$ 
saturates  at its value for $\omega=\delta_{\xi}$. 
However, in the insulator $K(\omega)\sim (E_{0}/\delta_{\xi})^{\gamma}\,
\ln^{d-1}(\delta_{\xi}/\omega)$ increases 
 further upon $\omega$ decrease \cite{CueKrav, MFSC}.  This logarithmic 
enhancement is due to the Mott's pairs of resonant levels which results in a 
well known \cite{Mott, Berezinskii} logarithmic enhancement of 
frequency-dependent conductivity $\sigma(\omega)\sim \omega^{2}\,\ln^{d+1}(\omega)$
 in insulator. The difference in the power of logarithm in 
$K(\omega)$ and $\sigma(\omega)$ is due to the square of the dipole 
moment matrix element entering  the conductivity. 
 Both limiting cases in a 3D insulator can be combined in one 
interpolating expression \cite{MFSC}:
\be
\label{interpolating}
K(\omega)=\frac{(E_{0}/\delta_{\xi})^{\gamma}\,\ln^{2}(\delta_{\xi}/\omega)}
{c+(\omega/\delta_{\xi})^{\gamma}\,\ln^{2}(\delta_{\xi}/\omega)},\;\;\;\;(c\sim 1).
\ee

\section{ Enhancement of cooling in a weak insulator.}
\begin{figure}
\center{\includegraphics[width=80mm]{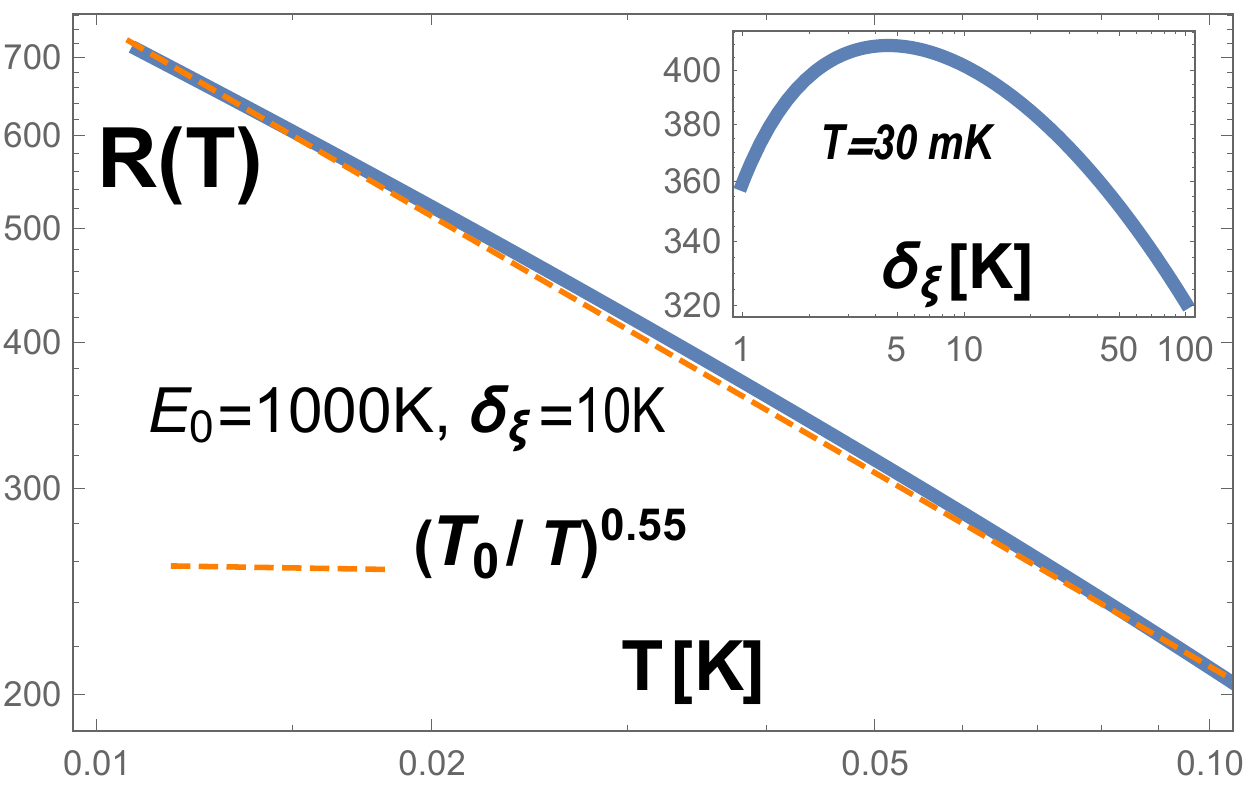}}
\caption{Enhancement factor $R(T)$ in Eqs. (\ref{JT}),(\ref{cool-longitud}) as a 
function of temperature for $E_{0}=1000\,K$ and $\delta_{\xi}=10\,K$. The dashed 
line represents a power law $(T_{0}/T)^{0.55}$ with $T_{0}=1710\,K$. Both the 
value of $R(T)$ and the exponent in the apparent power-law dependence are in 
agreement with experiment Refs.\cite{Ovadia-Sacepe-Shahar, Sacepe2018}. 
In the inset: the   enhancement factor $R$ as a function of $\delta_{\xi}$ for 
$T_{{\rm el}}=30\,mK$, $E_{0}=1000\, K$.}
\label{Fig:R}
\end{figure}

Because of the strong dependence of the cooling power $J\propto v_{s}^{-5}$ on 
the sound velocity $v_{s}$, the cooling is usually dominated by the transverse 
phonons which  sound velocity is typically smaller by a factor of about 2. 
Then neglecting the contribution of longitudinal phonons to cooling one obtains 
from (\ref{J-tau}):
\be
\label{JT}
J_{t}(T_{{\rm el}})=\frac{8}{5\pi^{2}}\frac{(k_{F}\ell)\,n_{{\rm e}}}{\rho_{i}\,
[v_{s}^{(t)}]^{5}}\,T_{{\rm el}}^{6}\,R(T_{{\rm el}}),
\ee 
where $T_{{\rm el}}$ is the temperature of electron system and the function
\be
\label{RT}
R(T)=\int_{0}^{\infty}dx\,x^{5}\,(\coth(x)-1)\,K(2T\,x).
\ee
Actually the integral in (\ref{RT}) is strongly peaked at $2x \approx 5$, thus 
the ratio $J(T_{el})/T^6_{el}$ is
proportional to  $K(5 T_{el})$.
In a limited interval of electron system temperatures $T_{{\rm el}}=10-100\,mK$ 
the enhancement factor $R(T_{{\rm el}})$ for typical parameters of Indium Oxide 
films $E_{0}=1000\,K$, $\delta_{\xi}=10\,K$, $c=1$  is well approximated by the 
power law $R(T)\approx \left(T_0/ T_{\rm el}\right)^{0.55}$ with
 $T_0 \approx 1700\,K$, see Fig.\ref{Fig:R}.
 Thus the effective power of temperature in $J(T_{{\rm el}})$ should be 
$\beta_{{\rm eff}}\approx 5.5$ rather than 6.0, in agreement 
with Ref.\cite{Sacepe2018}. The overall enhancement factor for
 this values of parameters varies from 700 to 200 at $T_{{\rm el}}=10\,mK-100\,mK$ 
which is consistent with experiment \cite{Ovadia-Sacepe-Shahar}.  The dependence 
of the $R(T)$ factor on the 
local level  spacing $\delta_{\xi}$ is rather weak, see inset to Fig.\ref{Fig:R}.

\section{ Cooling by longitudinal phonons.}
Considering the contribution of longitudinal phonons to   cooling rate, one has 
to take into account screening given by the second term in Eq.(\ref{screened}).
 The simplest case is the universal limit of screening when $V(q)\Pi\gg  1$ 
which is always the case in a 3D {\it metal} in the limit $q\rightarrow 0$ due 
to long-ranged Coulomb interaction $V(q)\propto 1/q^{2}$. In Anderson insulator 
this limit is approximate controlled by the  large value of the dielectric 
constant close to the localization transition \cite{IvanovCuevasFeigelman}.  
In this limit the electro-neutrality condition is strictly enforced and the 
second term in Eq.(\ref{screened}) takes the universal form
$-(\hat{{\cal D}}\Pi)\;\Pi^{-1}\;(\Pi \hat{{\cal D}})$.
One can approximate $\hat{{\cal D}}\Pi \approx \frac{\nabla{\bf u}}{m}\,k_{F}^{2}
\,\delta_{\alpha\beta}\,\Pi$, and $\Pi\hat{{\cal D}}\approx \frac{\nabla{\bf u}}{m}
\,k_{F}^{2}\,\delta_{\gamma\delta}\,\Pi $.  
Now proceeding in the same way as above using (\ref{psi-ansatz}),(\ref{a-corr}) 
and taking into account also the longitudinal part of (\ref{tau-fin}) we obtain 
the contribution of the longitudinal phonons to electron cooling:
\be\label{cool-longitud}
J_{l}(T_{{\rm el}})=\frac{24}{5\pi^{2}}\frac{(k_{F}\ell)\,n_{{\rm e}}}{\rho_{i}\,
[v_{s}^{(l)}]^{5}}\,T_{{\rm el}}^{6}\,R(T_{{\rm el}}).
\ee 
As in Eq.(\ref{JT}), this result differs only by a factor $R(T_{{\rm el}})$ from 
that for a disordered metal \cite{schmid, Reyzer}.

Note that the above method of calculation using the ansatz (\ref{a-corr}) 
is valid only for local contributions, as it completely ignores a possibility 
of building a  density-density propagator, the 'diffuson'. However, in the 
universal limit of screening the diffuson cannot be excited, as it is forbidden 
by electro-neutrality.   The effect of incomplete screening on the longitudinal 
phonon decay rate and cooling is much more involved ( see e.g. Ref.~\cite{Shtyk}).  It may play some
role in low-dimensional cases where the effects of incomplete Coulomb screening 
are stronger.
 
\section{ Conclusions.}
The main result of this paper is given by Eqs.(\ref{tau-fin},\ref{K-omega})  
which relates phonon decay rate
$1/\tau(\omega)$   due to inelastic interaction with electrons,  and  correlation 
function of the local 
density of states $K(\omega)$ characterizing  electron wave-functions near 
Anderson mobility edge. 
 A direct consequence of this relation is a strong enhancement of the 
electron-phonon 
cooling power in weak insulators, in comparison with usual diffusive metals, as 
demonstrated by Eqs.(\ref{JT},\ref{RT}) and Figs.\ref{Fig:enhance},\ref{Fig:R}.  
For the case of insulating 
Indium-Oxide films, studied in
Ref.\cite{Ovadia-Sacepe-Shahar}, this enhancement is estimated to be in the range 
of 500-1000,
in agreement with the experimental data.  In general, our results suggest that  
measurements of the
cooling rate Eq.(\ref{JT}) or ultrasound attenuation rate Eq.(\ref{tau-fin}) 
provide a direct access to the electronic 
local density of states (LDoS) correlation function $K(\omega)$.

On a more technical side, we expect that the same relation (\ref{tau-fin})
can be obtained by means of functional "sigma-model" approach like the one 
developed in~\cite{KamGlaz}.
 
The above results are general and valid for any 3D Anderson insulator with 
long localization length and
relatively weak Coulomb interaction (slight modification of our formulas will 
also work for 2D Anderson insulators).
In particular, one can use this approach to analyze the data on bistability of 
I-V characteristics and switching 
between high-resistance and low-resistance branches as function of applied 
voltage, as reported for a number
of various semiconductors or insulators, see  Refs.~\cite{1,2,3,4}.  However, 
one should keep in mind that
in insulators with strong Coulomb interaction it might be difficult to 
disentangle Coulomb correlation effects 
from purely localization effects.  In such a case effective correlation 
function $K_{\rm eff}(\omega)$ may
differ from its non-interacting version  given in  Eq.(\ref{interpolating}).

Our results for the electron-phonon cooling power make it possible to establish 
conditions for the 
observation of
many-body localization transition in electronic insulators, predicted 
theoretically more than decade ago~\cite{BAA}
but did not yet observed. 
One of the crucial problems to be solved in this 
respect is to find an insulator
with an extremely low thermal coupling between electrons and phonons, yet 
with measurable electric conductance.
Our theory will be instrumental to solve this important issue.

The behavior of the cooling power very similar to our prediction has been 
recently seen in the
resistive state of  moderately disordered superconducting Indium Oxide
 films at strong magnetic field and low temperatures: see Sec. IV
of the Supplementary Information to Ref.~\cite{Sacepe2018}, where $J(T_{\rm el}) 
\propto T_{{\rm el}}^{5.5}$ 
was observed. An enhancement (compared to the prediction for dirty metals with 
$k_{F}\ell \sim 0.3$) by a factor 400-800  of cooling power {\it per carrier} 
in insulating $Nb_{x}Si_{1-x}$  can also be extracted from the results of 
Ref.\cite{NbSi}.

Finally, we note that the obtained results are not expected to hold for 
pseudo-gaped insulators
where single-electron DoS is strongly suppressed due to local pairing~\cite{MFSC}.  Indeed, 
electron-phonon cooling rate in insulating state of Indium-Oxide realized at 
relatively low magnetic field
is known~\cite{Sacepe-personal}  to be much lower (and follow much faster 
temperature dependence)
 than the high-field data reported in Ref.~\cite{Ovadia-Sacepe-Shahar}. 
 The reason for that difference is that strong magnetic field 
(above approximately 10 Tesla) destroys
 local pairs and makes electron spectrum gapless.

\section{Acknowledgments.}
We are grateful to Claire Marrache, Benjamin Sacepe,  
Zvi Ovadyahu and Dan Shahar for numerous discussions of experimental data. 
This research was partially supported by  the Skoltech NGP grant. V.E.K. is 
grateful to M. A. Skvortsov for illuminating discussions and hospitality at 
Skoltech where the major part of this work was done. M.V.F appreciates 
hospitality of Abdus Salam ICTP where the final part of this work was done.

\appendix
\section*{Appendices}

\section{General expression for phonon attenuation rate in terms of electron wave functions. }

In order to take the localized nature of electron wave functions into account we express the   phonon attenuation rate in terms of the exact electron eigenfunctions $\psi_{n}({\bf r})$ and eigenvalues $E_{n}$.
Using (2)-(3) of the main text one can express the   contribution to $\tau_{ph}$ from the first term of (4) as follows:
\begin{widetext}
\bea\label{tau-K} 
 \frac{1}{\tau_{{\rm ph}}^{(1)}} =\frac{\pi}{2}\,\frac{q_{\beta}q_{\delta}}{m^{2}}\,e_{\alpha}e_{\gamma}\,\frac{1}{\rho_{i}\omega }\, \frac{1}{{\cal V} }  \int d E\int dE'\int d^{d}{\bf r}\int d^{d}{\bf r'}\,e^{i{\bf q}({\bf r}-{\bf r'})}\,F_{E,E'}(\omega)\,   K_{\alpha\beta\gamma\delta} ({\bf r},{\bf r'};E,E' ),
\eea
where $e_{\alpha}$ is the $\alpha$ component the unit vector of phonon polarization, $q_{\alpha}$ is the component of the phonon wave vector ${\bf q}$ with $|{\bf q}|=q=\omega/v_{s}$,    ${\cal V}$ is the volume, $F_{EE'}(\omega)=  \left[\tanh\left(\frac{E'+\omega  }{2T}\right)-\tanh\left(\frac{E'  }{2T}\right)\right]\, \delta(E'-E+\omega)$ is the Fermi distribution factor and the function 
$K_{\alpha\beta\gamma\delta} ({\bf r},{\bf r'};E,E' ) $  is defined as
\begin{equation}
\label{Ka}
K_{\alpha\beta\gamma\delta} ({\bf r},{\bf r'};E,E' )= \left\langle\sum_{nm} (\partial_{\alpha}\psi^{*}_{m}({\bf r}))\,(\partial_{\beta}\psi_{n}({\bf r}))\,(\partial_{\gamma}\psi^{*}_{n}({\bf r'}))\,(\partial_{\delta}\psi_{m}({\bf r'}))\,\delta(E-E_{n})\,\delta(E'-E_{m})\right\rangle.
\end{equation}
 
\end{widetext}
In (\ref{Ka}) we denote disorder averaging by $\langle .\rangle$. After such an averaging $K_{\alpha\beta\gamma\delta} ({\bf r},{\bf r'};E,E' )=K_{\alpha\beta\gamma\delta} ({\bf R},\varepsilon)$ becomes a function of ${\bf r}-{\bf r'}={\bf R}$ and $E-E'=\varepsilon$ in the bulk of the spectrum. One can use the (approximate) translation invariance in the energy space and perform integration over $E'$:
\be\label{int-E}
\int dE'\,F_{E,E'}(\omega)=2\omega\,\delta(\varepsilon-\omega).
\ee
Now the general expression for  $\tau_{ph}^{(1)}$ takes the following  form:
\begin{equation}
\label{tau-gen-a}
\frac{1}{\tau_{{\rm ph}}^{(1)}} =\pi\,\frac{q_{\beta}q_{\delta}}{m^{2}}\,e_{\alpha}e_{\gamma}\,\frac{1}{\rho_{i} }\,   
\int d^{d}{\bf R} \,e^{i{\bf q}{\bf R }}\,   K_{\alpha\beta\gamma\delta} ({\bf R}, \omega),
\end{equation}

\section{Phonon attenuation rate in disordered metals. }
At $k_{F}\ell\gg 1$ when the effects of multifractality can be neglected the correlation function $K(\omega)\approx 1$ at $\omega\gg \Delta$. Here we consider this limit in order to show that our approach based on Eq.(11),(12) of the main text (in which $\phi_{n}(r)=1$) reproduces the well known results of Refs. [15,16] where the diagrammatic approach was adopted. 

We start by evaluating the angular integrals over unit vectors ${\bf s}$ and ${\bf s'}$ in Eq.(14) of the main text. The result should have the following form:
\be\label{deltaS}
I_{1}\,\delta_{\alpha\delta}\delta_{\beta \gamma}+I_{2}\,(\delta_{\alpha\beta}\delta_{ \gamma\delta}+\delta_{\alpha\gamma}\delta_{ \beta\delta}),
\ee
as the integrals do not contain any preferential direction.

 The quantities $I_{1}$ and $I_{2}$ can be found from the following equations:
\begin{widetext}
\bea
9I_{1}+6I_{2}&=& \int d^{3}{\bf R}\,e^{-R/\ell}\,\left|\int  \frac{d\Omega_{{\bf s}}}{4\pi}\, e^{ik_{F}{\bf s}{\bf R}}\,  \right|^{2}, \label{II-1}\\
3I_{1}+12I_{2}&=& \int d^{3}{\bf R}\,e^{-R/\ell}\, \int  \frac{d\Omega_{{\bf s}}}{4\pi}\int \frac{d\Omega_{{\bf s'}}}{4\pi}\, e^{ik_{F}({\bf s}-{\bf s'}){\bf R}}\,({\bf s}{\bf s}')^{2}\label{II-2}.
\eea
\end{widetext}
The integral under the absolute value sign in Eq.(\ref{II-1})  is nothing but the Friedel oscillation in 3D space:
\be\label{Friedel}
\Psi_{{\rm Fried}}=\left(\frac{\sin(k_{F}R)}{k_{F}R}\right).
\ee
 Thus the R.H.S. of Eq.(\ref{II-1})  reduces to:
\be
\int d^{3}{\bf R}\,\left(\frac{\sin(k_{F}R)}{k_{F}R}\right)^{2} e^{-R/\ell}=\frac{8\pi\,\ell^{3}}
{1+4(k_{F}\ell)^{2}}.
\ee
The double angular integral in Eq.(\ref{II-2}) can also be expressed in terms of  $\Psi_{{\rm Fried}}$ and its second derivative:
\begin{widetext}
\be
\frac{1}{2}\left(\frac{\sin y}{y} \right)^{2}\,e^{-y/d}+ \left(\frac{\sin y}{y} \right)\,\partial^{2}_{y}
\left(\frac{\sin y}{y} \right)\,e^{-y/d}+\frac{3}{2} \left[\partial^{2}_{y}
\left(\frac{\sin y}{y} \right) \right]^{2} \,e^{-y/d},
\ee
\end{widetext}
where $y=k_{F}R$ and $d=k_{F}\ell$.

Now doing the ${\bf R}$-integral in Eq.(14) of the  main text we obtain:
\be
 \int d^{3}{\bf R}\,e^{-R/\ell}\, \int  \frac{d\Omega_{{\bf s}}}{4\pi}\int \frac{d\Omega_{{\bf s'}}}{4\pi}\, e^{ik_{F}({\bf s}-{\bf s'}){\bf R}}\,({\bf s}{\bf s}')^{2}=4\pi \ell^{3}\,Y(d),
\ee\label{d-all}
where the function $Y(d)$ is:
\be\label{d-all2}
Y(d)= \frac{2}{1+4d^{2}}+\frac{1}{d^{4}}-\frac{1+2d^{2}}{4d^{6}}\,\ln(1+4d^{2}).
\ee
In the limit $d=k_{F}\ell\gg 1$ one obtains from Eqs.(\ref{deltaS}),(\ref{II-1}),(\ref{II-2}),(\ref{d-all}):
\be\label{I-2}
I_{2}\approx I_{1}= \frac{2}{15}\,\pi\,\frac{\ell}{k_{F}^{2}}.
\ee
so that the combination of delta-symbols in Eq.(\ref{deltaS}) is totally symmetric.

 Now plugging this result into Eq.(14) of the main text one obtains the  transverse  phonon attenuation rate:  
\be\label{tau-trans}
\frac{1}{\tau^{(t)}_{{\rm ph}}}=\frac{q_{\perp}^{2}}{30\pi^{2}}\frac{k_{F}^{4}\ell}{\rho_{i}}=\frac{q_{\perp}^{2}}{10}\,
\frac{k_{F}\ell}{\rho_{i}}\,n_{e},
\ee
where $n_{e}$ is the total electron density (with both spin directions).

Correspondingly, the result for the out-cooling rate is:
\be
J_{t}(T)=\frac{4\pi^{4}}{630}\,\frac{1}{[v^{(t)}_{s}]^{5}\,\rho_{i}}\,(k_{F}\ell)\,n_{e}\,T^{6},
\ee
which coincides with the result of Refs.[15,16].

For longitudinal phonons Eq.(14) of the main text gives the result which is  by factor of 3 larger than in (\ref{tau-trans}):
\be\label{tau1-longitud}
\frac{1}{\tau^{(1,l)}_{{\rm ph}}}=\frac{q_{||}^{2}}{10\pi^{2}}\frac{k_{F}^{4}\ell}{\rho_{i}},
\ee
 However,  in order to compute the attenuation rate of longitudinal phonons one has to take into account also the second term in Eq.(4) of the main text. At a complete screening, this term has an opposite sign compared to (\ref{tau-trans}) and thus  the phonon attenuation rate for longitudinal phonons is smaller than in Eq.(\ref{tau1-longitud}).

The additional negative contribution  should be found from the expression similar to Eq.(7) of the main text:
 \be\label{tau-gen2}
\frac{1}{\tau_{{\rm ph}}^{(1)}} =\pi\,\frac{q_{\beta}q_{\delta}}{m^{2}}\,e_{\alpha}e_{\gamma}\,\frac{1}{\rho_{i} }\,   
\int d^{d}{\bf R} \,e^{i{\bf q}{\bf R }}\,   K^{(2)}_{\alpha\beta\gamma\delta} ({\bf R}, \omega),
\ee
where ${\bf R}={\bf r}-{\bf r'}$, $\omega=E-E'$, and the correlation function $K^{(2)}_{\alpha\beta\gamma\delta} ({\bf R}, \omega)$ is:
\begin{widetext}
\be\label{K2}
K^{(2)}_{\alpha\beta\gamma\delta} ({\bf R}, \omega)= - \frac{1}{9}\,\delta_{\alpha\beta}\delta_{\gamma\delta}\,k_{F}^{4}\,\left\langle\sum_{nm} \Psi^{*}_{m}({\bf r}) \, \Psi_{n}({\bf r}) \, \Psi^{*}_{n}({\bf r'}) \, \Psi_{m}({\bf r'}) \,\delta(E-E_{n})\,\delta(E'-E_{m})\right\rangle.
\ee
\end{widetext}
Now substituting Eq.(9),(10) of the main text into (\ref{K2}) we obtain in the limit $q\ell\ll 1$:
 \be\label{tau-2}
\frac{1}{\tau^{(2)}_{{\rm ph}}} =-\frac{q_{||}^{2}}{18\pi^{2}}\frac{k_{F}^{4}\ell}{\rho_{i}},
\ee
so that 
 \be\label{tau-longitud}
\frac{1}{\tau^{(l)}_{{\rm ph}}}=\frac{1}{\tau^{(1,l)}_{{\rm ph}}} +\frac{1}{\tau^{(2)}_{{\rm ph}}}=\frac{2q_{||}^{2}}{45\pi^{2}}\frac{k_{F}^{4}\ell}{\rho_{i}}.
\ee
Correspondingly, the out cooling rate due to longitudinal phonons is 
\be
J_{l}(T)=\frac{8\pi^{4}}{945}\,\frac{1}{[v^{(l)}_{s}]^{5}\,\rho_{i}}\,(k_{F}\ell)\,n_{e}\,T^{6}.
\ee
Thus the ratio of the  total contribution   of the transverse 
($2J_{t}(T)$) and the longitudinal ($J_{l}(T)$)  phonons to the cooling rate 
is $(3/2) [v_{s}^{(l)}/v_{s}^{(t)}]^{5}$, in agreement with earlier results 
( see e.g. Eq.(31) of Ref.[20] of the paper).

\end{document}